# Introducing a novel method for adaptive thresholding in brain tumor medical image segmentation

[1]Ali Fayzi Zengir , [2]Mohammad Fayzi, [3]Mostafa Forotan

[1] Machine Learning Engineer at AIFA (Artificial Intelligence Group of FANAP CO)

[2] Department of information management, Management Faculty, Tehran Branch, Islamic Azad University, Tehran, Iran

[3] Machine Learning Engineer at AIFA (Artificial Intelligence Group of FANAP CO)

[1] a.fayzi@fanap.ir
[2] MohammadFayzi@outlook.com
[3] m.forotan@fanap.ir

*Abstract*— One of the most significant challenges in the field of deep learning and medical image segmentation is to determine an appropriate threshold for classifying each pixel. This threshold is a value above which the model's output is considered to belong to a specific class. Manual thresholding based on personal experience is error-prone and time-consuming, particularly for complex problems such as medical images. Traditional methods for thresholding are not effective for determining the threshold value for such problems.

To tackle this challenge, automatic thresholding methods using deep learning have been proposed. However, the main issue with these methods is that they often determine the threshold value statically without considering changes in input data. Since input data can be dynamic and may change over time, threshold determination should be adaptive and consider input data and environmental conditions.

In this paper, we propose a novel method for adaptive thresholding in segmentation models using deep learning. This method can be applied to any segmentation model, increasing only the model parameters by the size of the input image. By considering environmental conditions and changes in input data, the proposed method adaptively determines the threshold value and can outperform traditional methods in complex problems with dynamic input data. In the following sections, we describe the details of this method and how it works, and present its results for medical image segmentation, particularly for brain tumor detection.

## I. INTRODUCTION

Detecting brain tumors is one of the major challenges in the field of medicine that still remains problematic due to its complexity and diversity in many cases. Accurate detection of brain tumors, especially in the early stages of the disease, can prevent serious side effects and sometimes the disabilities caused by surgical interventions. Therefore, the accurate and timely detection of brain tumors is of utmost importance.

In recent years, with the rapid advancement in the field of artificial intelligence, particularly in deep learning methods, many efforts have been made to use these methods in brain tumor detection. These methods, based on convolutional neural networks (CNNs), recurrent neural networks (RNNs), convolutional recurrent neural networks (CRNNs), and other deep neural networks, have significantly improved the accuracy of brain tumor detection.

In this paper, we propose a new method for adaptive threshold determination in brain tumor detection using deep learning. The adaptive threshold method is used to improve the accuracy and reduce the number of false detections in brain tumor detection. In this method, we use a deep neural network for brain tumor detection and then achieve a more accurate method for brain tumor detection by determining the adaptive threshold.

To evaluate the proposed method, we use a large dataset of brain images and brain tumors. The results show that the proposed method has a higher accuracy in brain tumor detection and significantly reduces the number of false detections. Moreover, this method is suitable in terms of time and can be used as a useful tool in brain tumor detection in various hospitals.

In the following sections, we first review brain tumor detection using medical images and deep learning methods. Then, we introduce the adaptive threshold and different methods of calculating it. Finally, we propose the proposed method for adaptive threshold determination in brain tumor detection using deep learning and examine the evaluation results.

Using medical images and deep learning methods is one of the important methods for brain tumor detection. Medical images of the brain include various types of images, such as radiology images, MRI and CT images of the brain and its tumors. The aim of using medical images is to diagnose and identify the location, size, detectability, and treatment of brain tumors.

One of the most common and effective methods for brain tumor detection is the use of deep learning methods. In this method, deep neural networks are used as a tool for brain tumor detection. Deep neural networks, due to their abilities in detecting complex patterns, have the best performance in detecting brain tumors from medical images.

Moreover, in recent years, deep neural networks have achieved higher accuracy in brain tumor detection using more advanced methods such as CNNs, RNNs, CRNNs, and other deep neural networks.

## II. RELATED WORKS

Detecting and treating brain tumors is one of the most important medical challenges. Medical brain images are used as a tool for diagnosing brain tumors. In this regard, the segmentation of medical brain images for the detection of brain tumors is of great importance.

Deep learning methods, such as convolutional neural networks, are used for the segmentation of medical brain images. U-Net is one of the commonly used convolutional neural networks for medical brain image segmentation.

In article [1], an improved adaptive segmentation algorithm is proposed, which can perform the most important process in building a 3D model, i.e., segmentation. The adaptive thresholding method can effectively detect the object boundary in an image intensity histogram, but tends to diverge in the smooth region. The proposed algorithm in article [1] examines the two sides of the histogram distribution and makes the adaptive thresholding algorithm more stable.

In article [2], a new adaptive thresholding technique for binarizing historical documents is proposed. The proposed method is a combination of Niblack and Sauvola techniques. By determining the mean threshold values of both methods, the drawbacks of Niblack and Sauvola are resolved.

Article [3] describes a local adaptive thresholding technique that removes the background using the mean and local mean deviation. The local computation time is usually dependent on the window size. The proposed technique in this article uses the integral image to calculate the local mean, which does not include standard deviation calculations like other local adaptive techniques.

In article [4], a supervised adaptive thresholding network is proposed for sample segmentation. Specifically, the Mask R-CNN method is adopted based on adaptive thresholding, and a layer-wise adaptive network structure is created to perform the dual adaptive thresholding on the probability graph generated by Mask RCNN to obtain better segmentation effects and reduce error rates. At the same time, an adaptive feature repository is designed to make the transfer between different layers of the network more accurate and effective, reduce errors in the feature transfer process, and further improve the Mask method.

Article [5] proposes a local adaptive thresholding technique based on energy information of gray-level co-occurrence matrix (GLCM) for retinal vessel segmentation. Different thresholds are calculated using the GLCM energy information. An empirical evaluation on the DRIVE database using the gray-scale intensity and green channel of the retinal image demonstrates the high performance of the proposed local adaptive thresholding method.

In article [6], an improved algorithm called Adaptive Thresholding by Gaussian Filtering and Adaptive Threshold Classification (ATC) is proposed for the segmentation of human X-ray images. The ATC technique is used to perform the thresholding process, which adapts to the image intensity distribution using a Gaussian filter.

The goal of article [7] is to automate the manual adjustment of thresholding in the PCNN edge detection model proposed in the previous stage. This is achieved by using the edge features extracted from the PCNN segmentation output and combining them with the DNN-based prediction to understand the adaptive thresholding parameter settings of the PCNN model. The Contourlet transform is used to extract feature vectors from the sub-bands of each iteration's output of the PCNN segmentation model. Then, median filtering is applied to the extracted feature vectors, followed by the calculation of variance and mean values as feature vectors. Finally, a DNN-based edge detection algorithm is proposed to achieve adaptive thresholding prediction. This method can achieve better edge detection results after only six iterations of the PCNN model.

In article [8], an adaptive thresholding method is studied and proposed for the segmentation of dermatoscopic images using the Gabor filter and principal component analysis (PCA). The Gabor filter is used to extract statistical image features, and PCA is used to transform these features into different bases. The proposed method is evaluated using the ISIC dataset, and the segmentation results are evaluated using the Dice and Jaccard similarity indices.

In article [9], two adaptive thresholding methods are proposed for the segmentation of skin lesions in dermatoscopic images. The proposed methods are developed based on the normalization of color channels using the NTSC and YCbCr color models.

## III. UNET ARCHITECTURE

The U-Net [10] is a special convolutional neural network used for medical image segmentation. It was introduced in 2015 and has many applications in medical image processing.

The U-Net architecture consists of two main parts: the encoder and the decoder. The encoder includes 2D convolutional and dropout layers that extract image features. The decoder includes combination and 2D convolutional layers that divide the image into segments and assign labels to different components of the image.

In the encoder part, the input image is transformed into various image features using 2D convolutional layers with different dimensions. Then, the different features are combined using combination layers with larger dimensions. Finally, the combined features are reduced to a feature map using 2D convolutional layers with larger dimensions.

In the decoder part, the feature map obtained from the encoder part is used. This feature map is divided into segments and assigned labels to different components of the image using combination layers with smaller dimensions and 2D convolutional layers with larger dimensions.

Due to its unique structure, the U-Net is highly effective in medical image segmentation tasks, particularly in the segmentation of medical images. Additionally, since the U-Net has a low number of parameters, it can be trained quickly and used in processing devices with limited resources. The U-Net model architecture is shown in Figure (1).

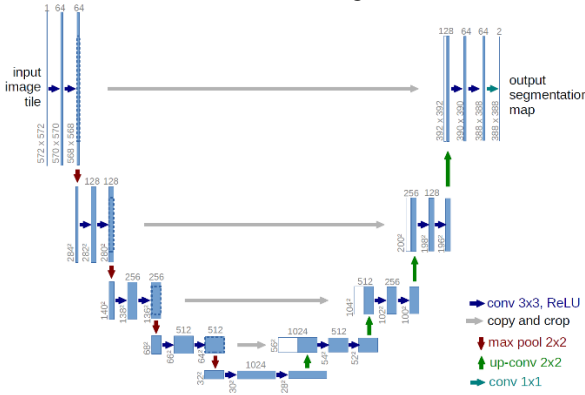

Figure (1): UNET model architecture

## IV. MOTIVATION

In existing models for image segmentation, a neural network is trained first, and then post-processing operations are performed on the network output, such as setting a threshold value to assign a value of 1 for output values above a certain threshold and 0 for the rest. However, this manual process increases the likelihood of misclassifying pixels. In this paper, an adaptive thresholding method is introduced that is learned automatically during the model training and applied to the model's segmentation output automatically. This method aims to reduce the error caused by manual thresholding and improve the accuracy of pixel segmentation.

## V. PROPOSED METHOD

Manual thresholding is a common challenge in image segmentation models, especially in medical image analysis. The process of manually selecting a threshold value for each pixel in an image is time-consuming and prone to errors. Different threshold values can lead to different segmentation results, making it difficult to compare the performance of different models. Moreover, the manual selection of threshold values is subjective and varies among different experts. This can lead to inconsistency in the interpretation of the results and limit the reproducibility of the analysis.

In the proposed method, a case study was selected as a test for the proposed method, in which the Unet model was used for brain tumor segmentation in medical images. A public dataset of brain tumor images was used in this paper, which includes MRI images of the human brain and a mask image of the brain tumor location. 80% of the data was used for Unet model training, 10% for model evaluation, and 10% for final model testing.

To evaluate the proposed method's accuracy improvement, we first trained the Unet network with the available data without any modifications. Then, we trained the Unet network with modifications, including the addition of an adaptive thresholding parameter, and reported the thresholding parameter's addition method and its results.

To obtain the final segmentation results from the trained models, a post-processing step is applied to the network output, as shown in Equation (1):

$$f(output) = \begin{cases} 0, & output < 0.5 \\ 1, & output \geq 0.5 \end{cases} \qquad (1$$

In this work, a trainable parameter is added to the network's parameters, which is equal to the size of the input image. This feature enables segmentation models to assign each pixel of the input image to the desired class based on the defined threshold. Figure (2) shows the Unet model structure with an added Adaptive Thresholding module.

The Adaptive Thresholding module is trained simultaneously with the network and adjusts the threshold value of the segmentation output based on the input image's characteristics. This module improves the accuracy of pixel segmentation and reduces the manual error caused by thresholding.

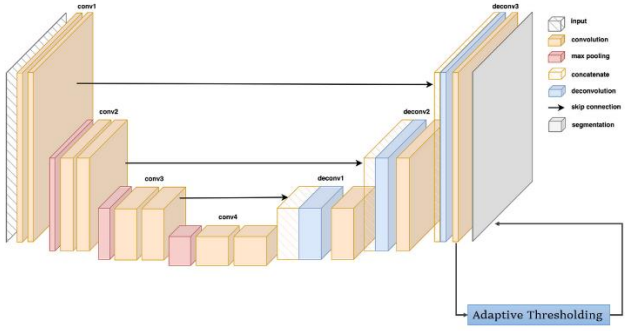

Figure (2): Structure of UNET model by adding Adaptive Thresholding

The structure of the Adaptive Thresholding module is shown in Figure (3).

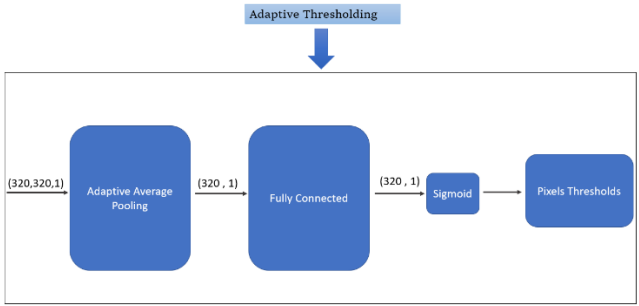

Figure (3): Adaptive Thresholding module architecture

In this module, the output of the Unet network, which has the same size as the input image but with a single channel, is first fed as input. Then, Adaptive Average Pooling is applied to it, and the output is fed into a Fully Connected layer, followed by a Sigmoid function. The desired threshold values for each pixel are obtained from the final output.

In this paper, the Simple Unet model was used to compare the performance of the proposed system with the addition of the Adaptive Thresholding module.

For training the Unet network for brain tumor segmentation in medical images, the Dice loss function (Equation (2)) is used to calculate the network's normal error, and the MSE loss function (Equation (3)) is used to calculate the error of the Adaptive Thresholding module. The Adam optimizer with a learning rate of 0.00005 is used for training.

$$DiceLoss = \frac{2*\sum p_{true}*p_{pred}}{\sum P_{true}^2 + \sum P_{pred}^2 + \epsilon} \qquad (2$$

$$MseLoss = \frac{1}{n}\sum(y_{true} - y_{pred})^2 \qquad (3$$

Python programming language and PyTorch framework were used to implement this paper, and the training was performed on a Linux operating system with an Nvidia RTX 1080 graphics card with 12GB memory.

The output results of the proposed system are shown in Figure (4).

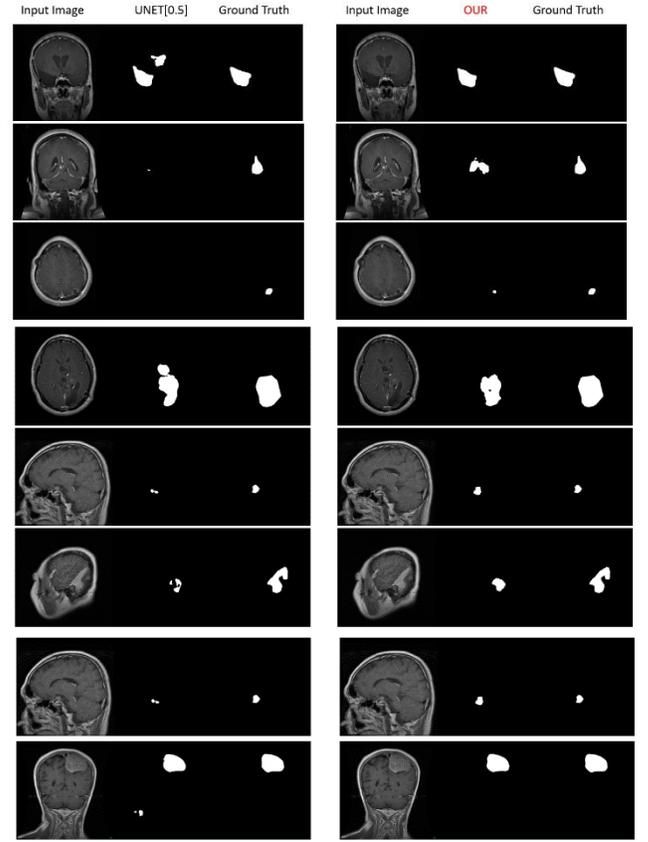

Figure (4): Comparison of Unet network outputs with a fixed threshold of 0.5 and the proposed Adaptive Thresholding method.

In this paper, a new method is proposed to solve the problem of adaptive thresholding in segmentation networks. By using this method, the challenges that arise in determining the adaptive threshold in image segmentation using conventional methods have been addressed.

In this method, a convolutional neural network with a specific architecture called U-Net is used as the main network for image segmentation. Using this network, the problem of adaptive thresholding is solved using an adaptive module. This module automatically determines the adaptive threshold using the image features and the threshold determined by the U-Net network.

The results obtained show that this method has a better accuracy in image segmentation compared to conventional adaptive thresholding methods. Additionally, due to the absence of the need for manual threshold determination, this method is faster in terms of time compared to conventional methods.

Overall, this new method for solving the problem of adaptive thresholding in segmentation networks using U-Net and the adaptive module provides positive results in medical image segmentation. This method can be used in many areas of medical image processing and can significantly improve the accuracy of medical image segmentation.